\documentclass{article}
\usepackage{spconf,amsmath,graphicx,hyperref,booktabs,multirow,makecell}
\usepackage{subcaption}
\usepackage{booktabs}
\usepackage{xcolor}
\title{TTA: Transcribe, Translate and Alignment \\ for Cross-lingual Speech Representation}
\name{Wei Liu$^{\dagger}$, Jiahong Li$^{\dagger}$, Yiwen Shao, Dong Yu \thanks{$^{\dagger}$Equal contribution.}}
\address{Tencent AI Lab, USA}
\begin{document}
\ninept
\maketitle
\begin{abstract}
% Speech-LLM models have demonstrated great performance in multi-modal understanding through speech modality integration within large language models (LLM).
Speech-LLM models have demonstrated great performance in multi-modal and multi-task speech understanding. 
A typical speech-LLM paradigm is integrating speech modality with a large language model (LLM). 
While the Whisper encoder was frequently adopted in previous studies for speech input, it shows limitations regarding input format, model scale and semantic performance.
To this end, we propose a lightweight TTA (Transcribe, Translate and Alignment) model specialized in speech semantics for more effective LLM integration. 
With large-scale training of 358k hours of speech data on multilingual speech recognition (MASR), speech translation (ST) and speech-text alignment tasks, TTA is capable of producing robust cross-lingual speech representations. 
Extensive evaluations across diverse benchmarks, including MASR/ST, speech retrieval, and ASR-LLM performance assessments, demonstrate TTA's superiority over Whisper. 
Furthermore, we rigorously validate the interplay between cross-lingual capabilities and MASR/ST performance.
The model weights \footnote{\scriptsize{https://huggingface.co/AudenAI/auden-tta-m10}} 
and training recipes of TTA will be released as part of an audio understanding toolkit \textit{Auden}
\footnote{\scriptsize{https://github.com/AudenAI/Auden/tree/main/examples/tta}}.
\end{abstract}
\section{Introduction}
\label{sec:intro}
 The remarkable success of large-scale pre-trained models in natural language processing has spurred significant interest in developing analogous speech foundation models. These models utilize a single architecture to perform a wide array of speech-related tasks to achieve generalized speech understanding. Multilingual speech recognition (MASR) and speech translation (ST) are two primary tasks to learn speech semantic information, such as Whisper~\cite{radford2023robust}, USM~\cite{zhang2023google}, OWSM~ \cite{DBLP:conf/asru/PengTYBCLSACSZSSJMW23}, and Canary~\cite{canary}.  More recently, the powerful Large Language Models (LLMs)~\cite{brown2020language,achiam2023gpt,team2024qwen2} in text comprehension and reasoning have led to efforts to integrate speech modality to develop speech-LLM~\cite{chu2023qwen,chu2024qwen2,DBLP:conf/iclr/TangYSC000M024, DBLP:conf/emnlp/HuZ0CMHPL0SLW24, lu2025desta2}, thereby enhancing multi-modal speech understanding. A common practice for speech-LLM is to attach a speech foundation model, often its encoder component, with an LLM to facilitate seamless cross-modal comprehension.
 
 Recent works on speech-LLM like Qwen-Audio~\cite{chu2023qwen,chu2024qwen2}, WavLLM \cite{DBLP:conf/emnlp/HuZ0CMHPL0SLW24}, SALMONN~\cite{DBLP:conf/iclr/TangYSC000M024}, DESTA2.5~\cite{lu2025desta2} mostly connect Whisper encoder to LLM. Qwen-Audio, for instance, pairs Whisper Large-v2 encoder with Qwen-7B~\cite{team2024qwen2} and achieves promising results in audio question-answering and chat tasks.  However, the Whisper encoder has several limitations standing out:
 (1) default constraint of 30-second speech input (unless fine-tuned); (2) focusing on speech semantics (with weaker performance for Chinese) while lacking paralinguistic information; and (3) large model sizes. To address these, we posit that a lightweight design can balance superior efficiency and performance. This paper focuses on developing a specialized speech semantic model. Unlike Whisper and its open-source reproduction OWSM, which primarily utilize a Transformer-based encoder-decoder architecture, ours employs a combined Zipformer-based~\cite{yao2023zipformer} Transducer (ZT) and attention-based encoder-decoder (AED) architecture. The choice of Zipformer stems from efficiency. This hybrid design (ZT-AED) is specifically engineered to enhance the semantic representational capacity of the encoder component. 

The training tasks used to learn speech semantics are MASR and ST. Prior work has empirically demonstrated the benefits of joint MASR-ST training~\cite{radford2023robust, DBLP:conf/asru/PengTYBCLSACSZSSJMW23, canary}. The basic assumption is that language-invariant information (e.g., semantics) can be shared. Learning cross-lingual speech representations aligned in a shared multilingual space is necessary. 
However, existing work has not clarified the exact relationship between cross-lingual representation and MASR/ST learning. These metrics may not always improve in tandem and the extent to which ST impacts MASR performance under fair settings all require further exploration.
% However, two key questions remain unaddressed: the under-explored relationship between cross-lingual representation and MASR/ST learning, and the extent to which ST impacts MASR performance under fair settings. 
% (e.g., identical speech data sources).
% However, the relationship between cross-lingual representation and ASR/ST learning tasks has not been fully explored. Also, to what extent the ST would impact the ASR performance under a fair experimental setting (e.g., using the same speech data source) remains unclear.

In this study, we introduce \textbf{TTA} (\textbf{T}ranscribe, \textbf{T}ranslate and \textbf{A}lignment), a highly efficient speech semantic foundation model, less than 250M parameters. TTA employs an innovative ZT-AED joint architecture optimized to capture rich speech semantics via MASR and ST learning. 
%In addition, a contrastive loss is introduced to align speech features and text embeddings in a multilingual space. This semantic alignment process explicitly helps learn cross-lingual representations. With around 250M parameters, TTA is trained on 358 thousand hours of speech data and outperforms Whisper-medium (769M) in both MASR and ST benchmarks.
It further incorporates a contrastive loss to align speech features and text embeddings in a multilingual space, explicitly enhancing cross-lingual representation learning. Trained on 358k hours of speech data, TTA consistently outperforms Whisper Medium on multiple MASR and ST benchmarks, while surpassing Whisper Large in some in-domain benchmarks. %  (769M parameters)
Our key contributions are: (1) A lightweight model exceeding Whisper-medium performance, validating the ZT-AED architecture’s effectiveness; (2) Enhanced cross-lingual speech representations via alignment loss, significantly boosting speech retrieval performance. Furthermore, we provide insights into the interplay between cross-lingual capability, MASR, and ST; (3) Demonstrated speech semantic superiority of TTA’s encoder over Whisper encoders for LLM integration.
% Then, several key contributions of this work are listed as follows: (1) We provide a lightweight speech semantic model with performance surpassing Whisper-medium, and some languages beat Whisper-large, meanwhile validating the effectiveness of ZT-AED architecture. (2) We leverage an alignment loss to boost speech representation's cross-lingual ability and explore the relationship among cross-lingual, ASR, and ST. Furthermore, our TTA model exhibits significantly better speech retrieval performance thanks to the cross-lingual enhancement. (3) We examine the superiority of the TTA model's encoder to connect the LLM compared to the commonly used Whisper encoders.

\section{TTA: Transcribe, Translate and Alignment}
\begin{figure}[h!]
    \vspace{-5mm}
  \centering
  \includegraphics[width=0.8\linewidth, height=5.5cm]{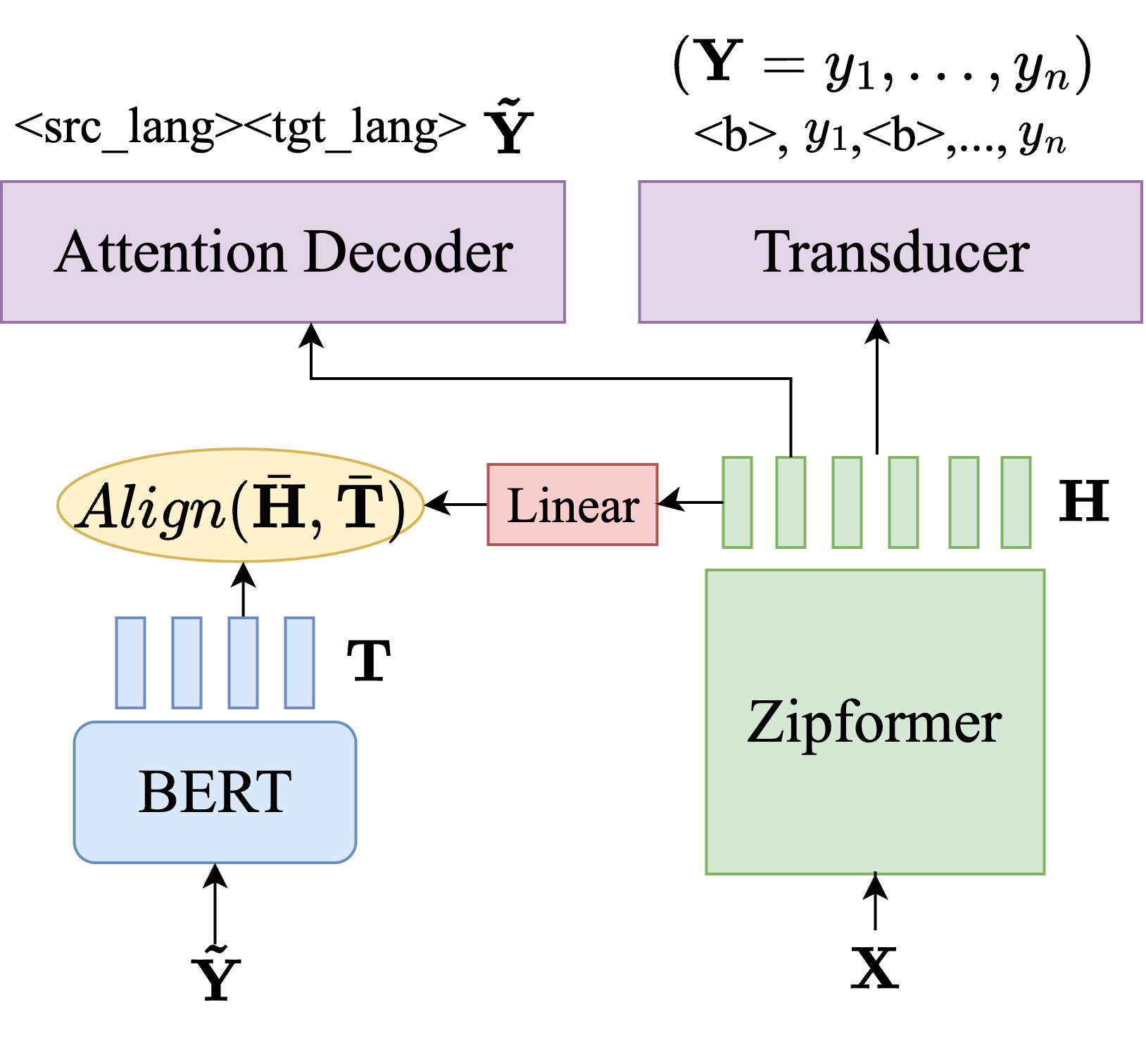}
\caption{The overall diagram of TTA model.}
\label{fig:tta}
 \vspace{-3mm}
\end{figure}

TTA adopts a hybrid ZT-AED architecture as shown in Figure \ref{fig:tta}. 
The model integrates three components: a Zipformer-based Transducer, an attention-based Transformer decoder, and a BERT-based~\cite{devlin2019bert} speech-text alignment module. 
The Zipformer encoder is a fast and memory-efficient variant of the Conformer~\cite{gulati2020conformer} architecture, which serves as the backbone to encode the input speech feature $\mathbf{X}$ into high-level speech representations $\mathbf{H}$. 
These representations are subsequently processed by three distinct branches. 

The Transducer branch is composed of a decoder and a joiner network for ASR. 
The decoder processes non-blank text tokens while the joiner performs token predictions.
In Figure \ref{fig:tta}, \texttt{<b>} denotes the special blank token in Transducer and $\mathbf{Y} = \{y_i\}_{i=1}^{n}$ is the transcription corresponding to $\mathbf{X}$. 

In the attention decoder branch, $\mathbf{H}$ is used to autoregressively generate the token sequence $\mathbf{\tilde{Y}}$. 
To support MASR, ST and language identification (LID), two special tokens of \texttt{<src\_lang>} and \texttt{<tgt\_lang>} are incorporated to represent the source and target language. 
The \texttt{<src\_lang>} token is trained to predict the spoken language, and the $\texttt{<tgt\_lang>}$ token implicitly determines the generation task: `\texttt{transcribe}' when two tokens match, and `\texttt{translate}' when they differ. 
In the latter case, $\mathbf{\tilde{Y}}$ corresponds to the target language translation rather than the source transcription $\mathbf{Y}$, which is different from the Transducer branch.

The last branch is designed for speech-text semantic alignment in the multilingual embedding space. 
A frozen multilingual BERT encoder\footnote{\scriptsize https://huggingface.co/google-bert/bert-base-multilingual-uncased} is employed to provide the semantic anchor. Specifically, the BERT encoder extracts text embedding $\mathbf{T}$ from $\mathbf{\tilde{Y}}$. 
The speech representation $\mathbf{H}$ is projected by a linear network to match the dimension of $\mathbf{T}$, and a SigLIP~\cite{zhai2023sigmoid} contrastive loss is applied on the a batch of $\{\bar{\mathbf{H}}, \bar{\mathbf{T}}\}_{i}$, where $\bar{\mathbf{H}}$ and $\bar{\mathbf{T}}$ are the utterance-level average pooling of the sequence $\mathbf{H}$ and $\mathbf{T}$ respectively. 
This contrastive objective minimizes the distances between positive pairs and pushes other negative, semantically dissimilar pairs further apart.  

Overall, TTA is designed to integrate multilingual speech recognition, speech translation and speech-text alignment within a lightweight model. 
These tasks can equip the model with the capabilities of general cross-lingual speech understanding.

\begin{figure}[htb]
\vspace{-3mm}
  \centering
  \includegraphics[width=0.9\linewidth]{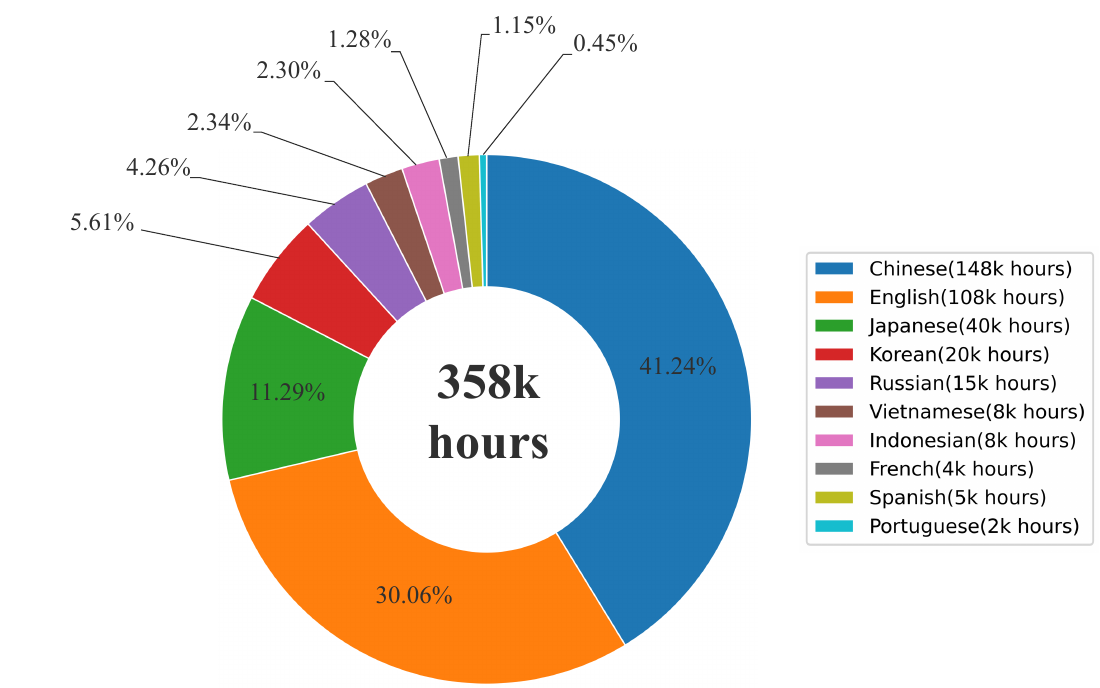}
\caption{Statistics of ASR training data over 10 languages.}
\label{fig:data_distribution}
\end{figure}

\begin{table*}[t!]
\vspace{-3mm}
\centering
\caption{Performance of multilingual speech recognition and speech translation. Four TTA series models are compared against Whisper series models, where ZT represents Zipformer-Transducer and AED denotes attention encoder decoder. ``(asr)" indicates that the models are trained exclusively on ASR data. TTA has only one extra alignment component vs. ZT-AED architecture.}
\label{tab:asrast-main}
 \resizebox{0.85\linewidth}{!}{
\setlength{\tabcolsep}{4pt}
\renewcommand{\arraystretch}{1.0}
\scalebox{0.4}{
\begin{tabular}{l|c|ccc|cccc}
\toprule
\multicolumn{1}{c|}{\multirow{2}{*}{\textbf{Datasets}}} & \multirow{1}{*}{\textbf{Metric}} & \textbf{Whisper} & \textbf{Whisper} & \textbf{Whisper} & \multirow{1}{*}{\textbf{ZT}} & \multirow{1}{*}{\textbf{ZT-AED}} & \multirow{2}{*}{\textbf{ZT-AED}} & \multirow{2}{*}{\textbf{TTA}} \\
~ & (\%) & {Medium} & {Large-v2} & {Large-v3} & (asr) & (asr) & ~  & ~ \\
\midrule
\multicolumn{2}{c|}{\#Params} & 762M & 1541M & 1542M & 199M & 246M & 246M & 247M \\
\midrule
\textbf{aishell} 1$|$2 & \multirow{2}{*}{{CER$\downarrow$}} & 6.74$|$6.23 & 5.90$|$5.24 & 5.33$|$4.76 & 1.89$|$3.14 & 1.82$|$3.07 & 1.80$|$3.03 & 1.85$|$3.09 \\
\textbf{wenet} net$|$meeting & ~ & 11.00$|$22.68 & 9.47$|$22.77 & 9.00$|$15.68 & 6.91$|$6.08 & 6.89$|$6.18 & 6.96$|$5.94 & 7.06$|$6.44 \\
\midrule
\textbf{librispeech} clean$|$other & \multirow{3}{*}{{WER$\downarrow$}} & 2.88$|$6.08 & 2.64$|$5.14 & 2.01$|$3.89 & 1.58$|$3.62 & 1.54$|$3.59 & 1.56$|$3.76 & 1.58$|$3.85 \\
 \textbf{gigaspeech} & ~ & 15.51 & 15.75 & 14.53 & 14.85 & 14.76 & 14.99 & 14.97 \\
\textbf{AMI} & ~ & 16.77 & 17.07 & 15.98 & 11.11 & 10.85 & 10.76 & 11.06 \\
\midrule
\textbf{commonvoice} & \multirow{4}{*}{\makecell{WER\\avg$\downarrow$}} & 11.86 & 9.70 & 8.30 & 6.92 & 6.70 & 6.69 & 6.76 \\
\textbf{MLS} & ~ & 7.27 & 5.65 & 4.48 & 5.82 & 5.71 & 5.72 & 5.74 \\
\textbf{voxpopuli} & ~ & 12.08 & 11.90 & 13.78 & 11.12 & 10.78 & 10.88 & 10.87 \\
\textbf{fleurs} & ~ & 6.62 & 5.20 & 4.51 & 6.35 & 6.18 & 6.17 & 6.19 \\
\midrule
\textbf{covostv2} & {BLEU$\uparrow$} & 35.12 & 38.80 & 37.60 & - & - & 34.72 & 35.28 \\
\bottomrule
\end{tabular}
}
}
\end{table*}

\section{Setup}

\subsection{Data Collection}
% asr data
\noindent\textbf{ASR training data.} 
The training data comprise a combination of in-house and open-source datasets covering 10 languages: Chinese (zh), English (en), Japanese (ja), Korean (ko), Russian (ru), Vietnamese (vi), Indonesian (id), French (fr), Spanish (es) and Portuguese (pt). 
Figure.\ref{fig:data_distribution} illustrates the language distribution of the total 357,982 hours of ASR training data. 
Approximately half of this corpus was collected from publicly available datasets, such as Aishell~\cite{bu2017aishell,du2018aishell}, WenetSpeech~\cite{zhang2022wenetspeech}, LibriSpeech~\cite{librispeech}, AMI~\cite{ami}, MLS~\cite{pratap2020mls}, VoxPopuli~\cite{wang2021voxpopuli}, LibriHeavy~\cite{kang2024libriheavy}, GigaSpeech~\cite{chen2021gigaspeech}, CommonVoice~\cite{ardila2020common}.
To ensure high data quality, all the data sources have undergone rigorous filtering procedures.  
Specifically, Whisper Large-v3 was employed to identify and exclude samples with incorrect language labels, and verify transcription using a WER threshold of 10-20\%.
% (decided by Whisper's recognition performance on different languages). 

% ast data
\noindent\textbf{AST training data.} 
For speech translation, our supervised data are restricted solely to the X$\xrightarrow{}$EN splits from CoVoSTv2~\cite{wang2020covost} and Europarl-ST~\cite{iranzo2020europarl}.
Synthesized X$\xrightarrow{}$EN translation pairs were generated from the aforementioned ASR training data using 
LLM\footnote{\scriptsize https://huggingface.co/Qwen/Qwen2.5-7B-Instruct}, 
with heuristic rules to eliminate potential hallucinations.
The resulting ST data comprises approximately 217,000 hours in total, with each sample explicitly linked to its ASR data source.
This enables precise control over data sampling ratio between MASR and ST training.

\noindent\textbf{Testing data and decoding.} 
Test sets from public datasets were used to measure MASR/ST performance across all the 10 languages. 
For MASR, Aishell, WenetSpeech, LibriSpeech, GigaSpeech, and AMI were adopted for zh/en assessment; 
CommonVoice-15, MLS, and VoxPopuli were adopted for multilingual assessment;
Fleurs~\cite{conneau2023fleurs} was preserved for zero-shot testing with its training set excluded from our training data.
The translation capability was examined by test data of CoVoSTv2.
In MASR decoding, we use the transducer branch to perform language-agnostic greedy search. 
The recognition accuracy of attention decoding from the AED branch is slightly better than transducer in beam search, but similar in greedy search. We use it for LID and ST decoding.
The Whisper baselines adopt attention greedy search, providing the ground truth language.
WER/CER is used to evaluate MASR performance (CER is for CJK languages), while BLEU is used to score ST quality.
 
\noindent\textbf{Text normalization.} 
To fairly compare different systems, we applied strict text normalization procedures.
Following the practice of Whisper
\footnote{\scriptsize https://github.com/openai/whisper/tree/main/whisper/normalizers}, 
all special symbols (expect the apostrophe) are removed, and a rule-based English normalization is adopted.
An extra Chinese number normalization is applied following Qwen2-Audio
\footnote{\scriptsize {https://github.com/QwenLM/Qwen2-Audio/blob/main/eval\_audio/cn\_tn.py}}.
% All special symbols (expect the apostrophe) are removed. Rule-based English normalization is adopted following the practice of Whisper and ruled-based Chinese number normalization is adopted following Qwen2-Audio~\cite{chu2024qwen2}.
% Additionally, texts are normalized by `NFKC' and `NFKD'\footnote{Two Unicode standard for text normalization.} in the training and testing stage, respectively. 

\subsection{Training Settings}
TTA model is composed of a Zipformer-large encoder with output dimension of 256, a standard Transducer network~\cite{DBLP:conf/interspeech/KuangGKLLYP22}, a 6-layer attention decoder, and a contrastive module for speech-text alignment.
The encoder consumes filter-bank features from 80-dim log-mel spectrum extracted with a window of 25 ms and a stride of 10 ms.
A unigram \textit{SentencePiece}~\cite{kudo2018sentencepiece} tokenizer is trained on the training corpus with a vocabulary size of 19,763.
% The multilingual vocabulary size is 19,763 and the corresponding tokenizer is trained using \textit{SentencePiece}~\cite{kudo2018sentencepiece} (unigram type) in our corpus.
Data-loading details are implemented with the Lhotse~\cite{zelasko2021lhotse} framework, which features an efficient batching strategy with data buffers to shuffle and group samples according to their duration.
The Transducer loss is averaged with the attention decoder loss, and the alignment loss is added with the weight of 0.1.
Models are trained on 32 V100 GPUs with \texttt{DynamicBucketingSampler} of maximum 250 seconds, \texttt{Scaled\_Adam} optimizer with a peak learning rate of 0.035, and \texttt{Eden} scheduler with warm-up of 2000 steps.
To mitigate data imbalance over different languages and data sources, we adopt a  similar sampling policy as~\cite{canary}: the dataset muxing weights $w_i$ are modified to $w_i=h_i^t$, scaling the total duration $h_i$ of dataset $i$ with temperature $t$. 
% , \texttt{num\_buckets} of 10, \texttt{buffer\_size} of 200000 and \texttt{shuffle\_buffer\_size} of 50000.
To make the training process more stable, TTA was trained with a multi-stage startegy.
In Stage 1, a ZT model was trained on ASR data for 250,000 steps.
Next in Stage 2, different models were initialized from the same ZT checkpoint and continued the ASR training from lr of 0.005 for 200,000 steps.
In Stage 3, models with ST were trained on the joint ASR/ST data with a $3:2$ mixing ratio from lr of 0.002 for 500,000 steps, while the ASR-only models were kept the same. 
$t$ was originally set as $1.0$ and gradually adjust to $0.2$ during the last training stage.
% The models were trained with Tencent Auden framework with 32 V100(32GB) GPUs.

\section{Results}

\subsection{Performance Results on ASR, ST, and LID}
 A comprehensive evaluation of ASR and ST performance across multiple benchmarks is presented in Table \ref{tab:asrast-main}. 
 The ZT-based models are directly compared against the Whisper series in terms of model scale, recognition WER, and translation BLEU score. 
 On widely used Chinese and English test sets, including Aishell, WenetSpeech, and LibriSpeech, the TTA model significantly outperforms Whisper models. This performance advantage can be attributed in part to the substantial proportion of Chinese and English data within our training data.
 In Multilingual ASR benchmarks, the TTA model demonstrates a considerable WER on the CommonVoice dataset, even compared with Whisper Large-v3 (6.76\% vs. 8.30\%).
 Clear advantages on MLS and VoxPopuli datasets are also observed. 
 % We attribute this result to the benefit of in-domain training. 
 For zero-shot evaluation on Fleurs, the TTA model fails to surpass Whisper Large models, while still behaves better than Whisper Medium. 
 This partially examines the recognition generalization of the TTA model, given its lightweight design and significantly fewer parameters even compared to Whisper Medium. % less than Whisper medium's 762M model parameters 
 
For speech translation performance measured with BLEU score on CoVoSTv2, the TTA model exhibits better performance than Whisper Medium, while still lagging behind Whisper Large models. This ST performance ceiling is primarily constrained by model capacity. 
We validate this assumption on another model with doubled model dimension which shows a significant boost. % and beating Whisper Large-v3. 

To further validate the effectiveness of the TTA architecture, an ablation study is conducted on the probing task of ST.
Several models from training stage 2 without ST training are selected for comparison, namely ZT, ZT-AED, ZT-Align and TTA. 
Their encoders are frozen and connected to a randomly initialized attention decoder to train the ST task. 
Validation loss curves averaged over 10 language pairs of each model during ST training are plotted in Figure \ref{fig:training_dynamic} (a). 
We found that models with \textit{Alignment} components perform consistently better. 
This suggests that explicit semantic alignment through the speech-text alignment module provides effective multilingual semantic anchors, thereby enhancing the encoder's cross-lingual representations and facilitating speech translation.

TTA also demostrates strong capability on the LID task. 
As evaluated on Fleurs, it achieves 100\% accuracy for all 10 training languages. 
In contrast, Whisper Large-v3 performs generally the same but worse in Indonesian with 81\% accuracy.
% , although it supports more languages.

\subsection{Cross-lingual Speech Retrieval}
\label{sec:speech_retrieval}
Speech-to-speech retrieval provides a direct approach to assessing the cross-lingual capabilities of semantic representations.
Speech samples with the same semantic meaning in different languages are supposed to encoded closer in the embedding space. 
Following~\cite{DBLP:conf/naacl/MaQFTGK25}, the \texttt{dev} and \texttt{test} sets from Fleurs are merged as an evaluation corpus with approximately 500 semantically aligned samples across all languages. 
We extract speech representations using various encoders and performed cross-lingual retrieval based on cosine similarity, where each query utterance is matched against all candidate utterances from another language.

\begin{figure}[t]
 \centering
 % Subplot 0-0
 \begin{subfigure}[t]{0.3\linewidth}
 \centering
 \includegraphics[width=\linewidth]{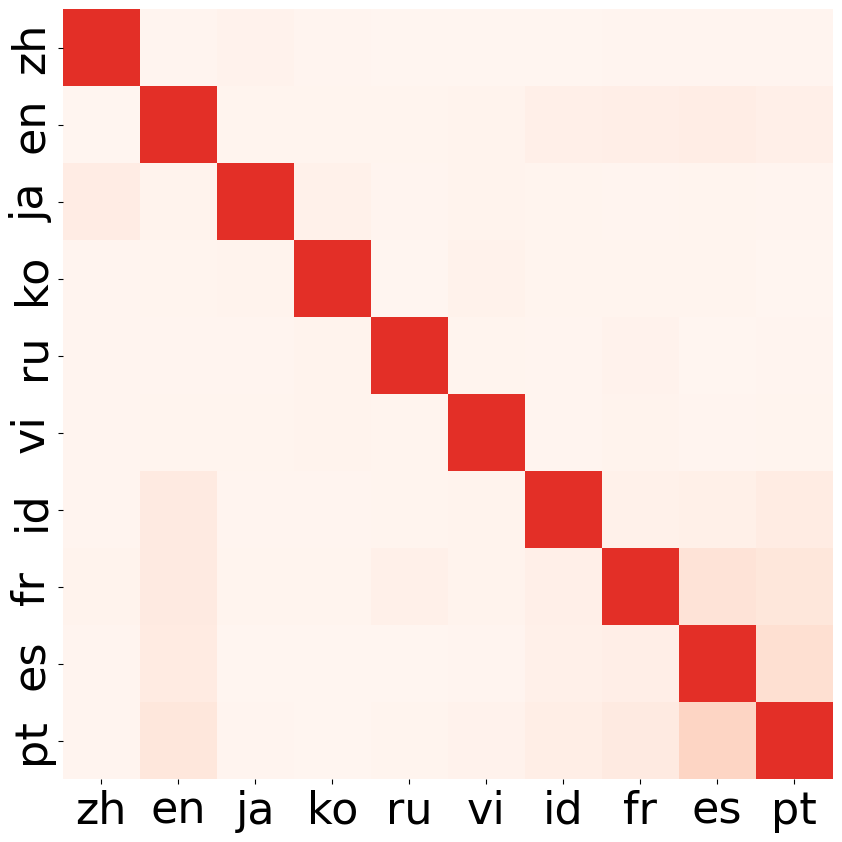}
 \caption{ZT(asr)}
 \end{subfigure}
 % Subplot 0-1
 \begin{subfigure}[t]{0.3\linewidth}
 \centering
 \includegraphics[width=\linewidth]{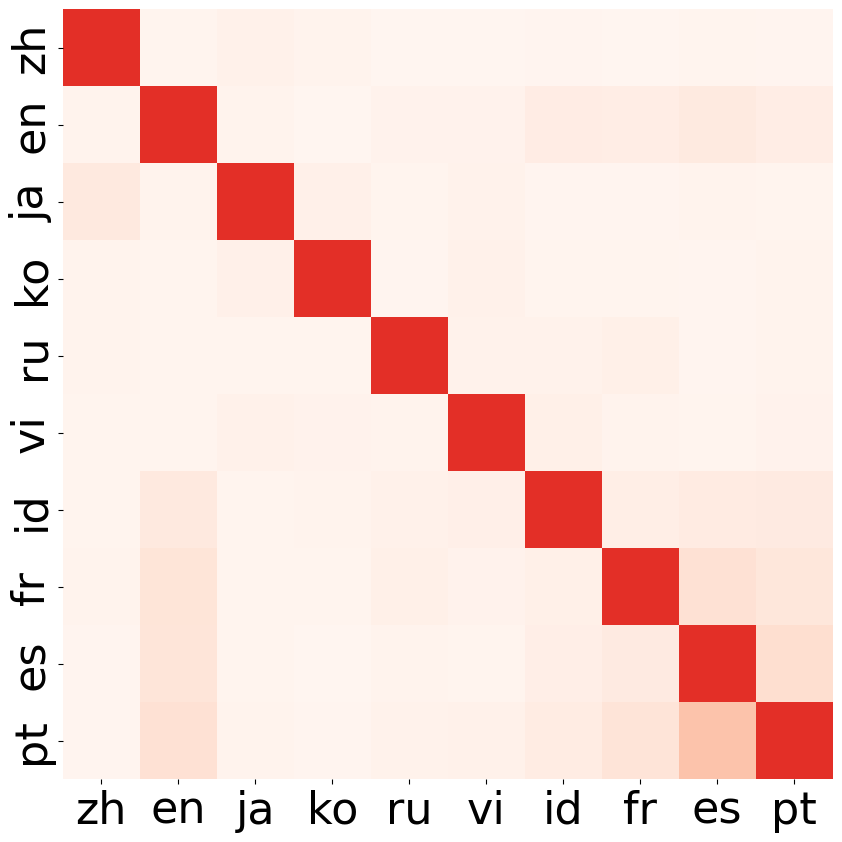}
 \caption{ZT-AED(asr)}
 \end{subfigure}
 % Subplot 0-2
 \begin{subfigure}[t]{0.3\linewidth}
 \centering
 \includegraphics[width=\linewidth]{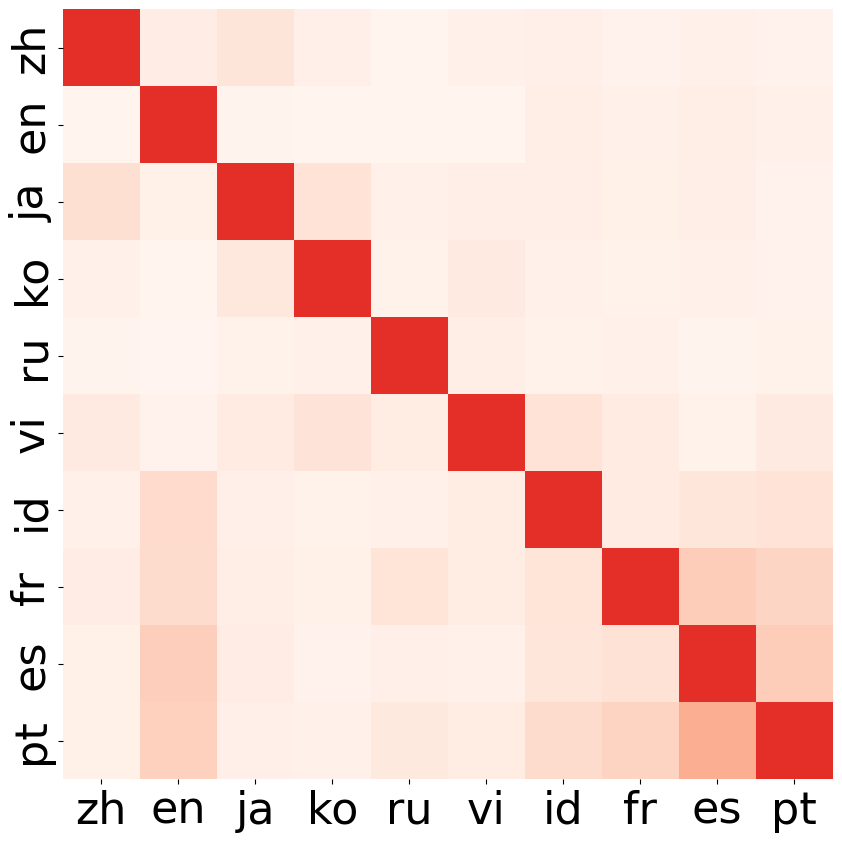}
 \caption{ZT-AED}
 \end{subfigure}
 % Subplot 1-0
 \begin{subfigure}[t]{0.3\linewidth}
 \centering
 \includegraphics[width=\linewidth]{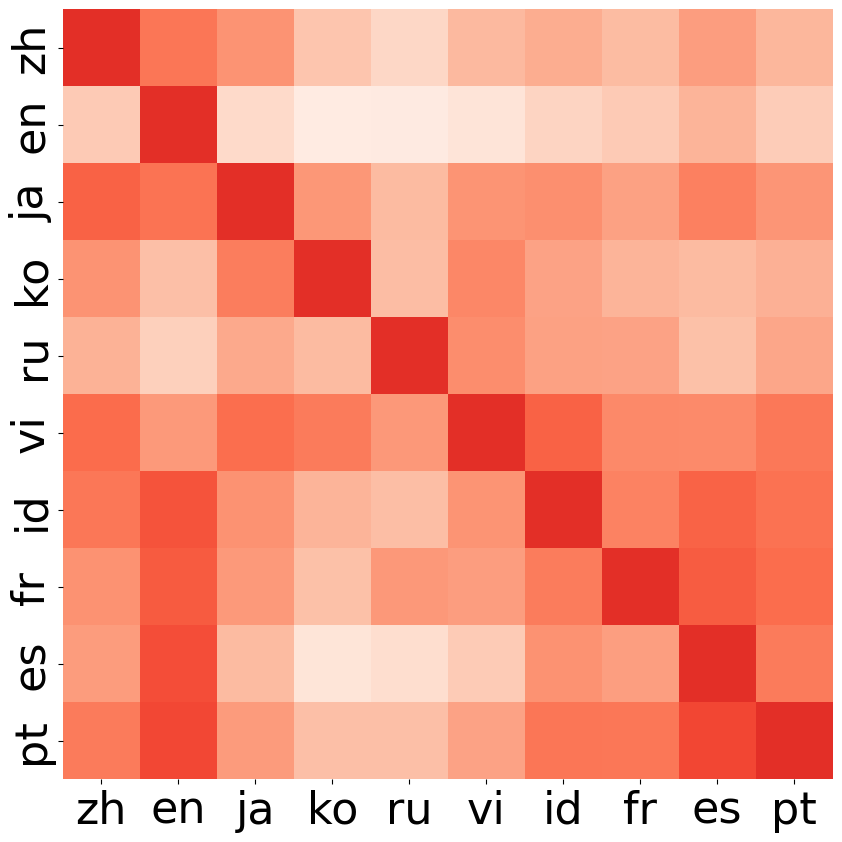}
 \caption{TTA}
 \end{subfigure}
 % Subplot 1-1
 \begin{subfigure}[t]{0.3\linewidth}
 \centering
 \includegraphics[width=\linewidth]{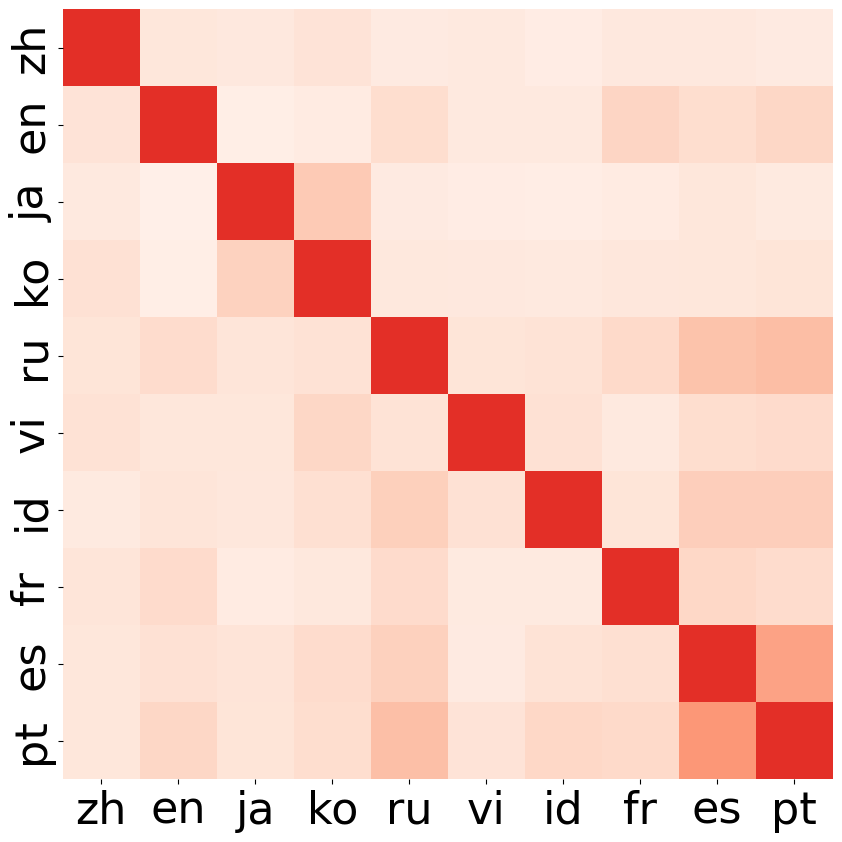}
 \caption{Whisper-M}
 \end{subfigure}
 % Subplot 1-2
 \begin{subfigure}[t]{0.3\linewidth}
 \centering
 \includegraphics[width=\linewidth]{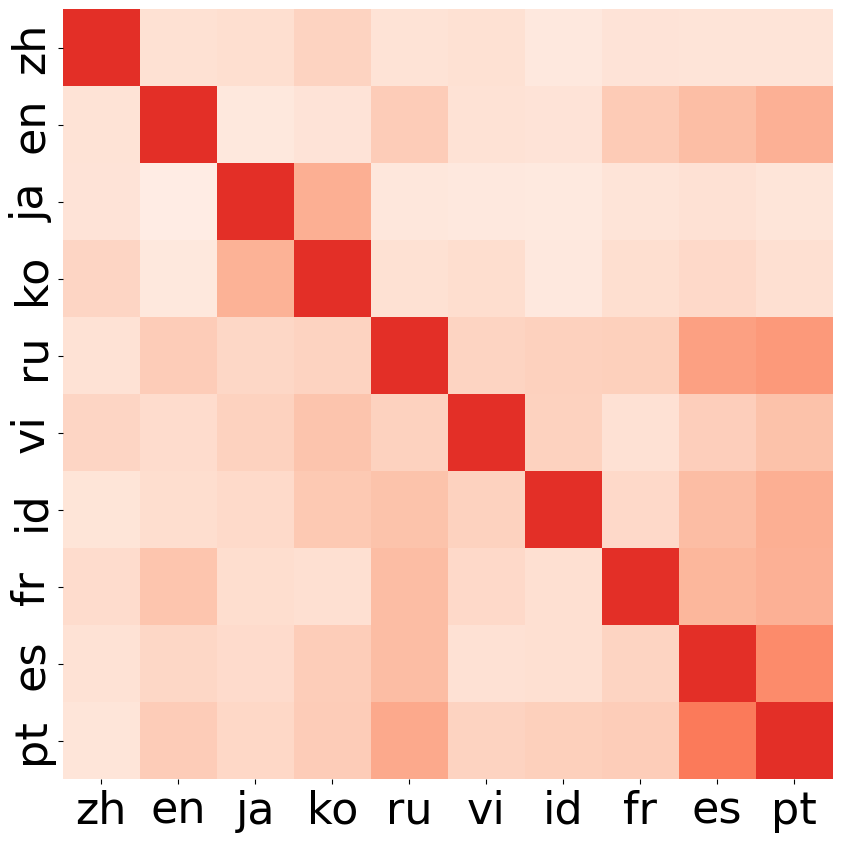}
 \caption{Whisper-L}
 \end{subfigure}
  
  \caption{Heatmaps of speech-to-speech retrieval accuracy across 10 languages. Inside each subfigure, each cell represent the retrieval accuracy from language X to language Y. The deeper color indicates higher retrieval accuracy and better cross-lingual alignment.} 
  \label{fig:speech_retrieval}
  \vspace{-4mm}
\end{figure}

The detailed results of speech retrieval are presented in Figure \ref{fig:speech_retrieval}. 
Retrieval accuracy is generally higher among Indo-European languages, which can be attributed to their linguistic proximity. 
In terms of model performance, the ZT-AED model, which incorporates speech translation, slightly outperforms ASR-only baselines and performs comparably to Whisper Medium. 
Notably, the TTA model achieves a substantial improvement in retrieval accuracy, even surpassing Whisper Large-v2. 
This enhancement underscores the efficacy of its explicit speech-text alignment mechanism in learning language-agnostic semantic representations.

\subsection{Relationship between Cross-lingual and MASR/ST}
We further inspect Table \ref{tab:asrast-main} by comparing four TTA series models under identical conditions, i.e., with the same training steps and learning schedule. 
These model includes: ZT(asr), ZT-AED(asr), ZT-AED, and TTA. 
Models marked with ``(asr)" are trained exclusively on ASR data, and ZT-AED differs from TTA by a single \textit{Alignment} component.

The performance gap between ZT(asr) and ZT-AED(asr) highlights the advantage of the hybrid architecture. 
Comparing TTA with ZT-AED, the additional cross-lingual alignment yields clear benefits for ST performance, with increase of approximately 0.6 BLEU score. 
However, the \textit{Alignment} is observed to bring minor degradation in ASR performance. 
With careful tuning of the \textit{Alignment} weight, this degradation of less than 0.1\% WER can be neglected. 

On the other hand, it implies that ASR performance may not really benefit from the cross-lingual ability enhancement. 
% In a normal multilingual ASR training, models spontaneously learn and share the language-invariant information across multiple languages. 
As indicated in Figure \ref{fig:speech_retrieval}, ZT-AED(asr) has shown basic cross-lingual abilities when learning and sharing the language-invariant information across multiple languages. For example, it tends to align the language \textit{en} together with \textit{\{fr,es,pt\}} languages. 
Enforcing a more aggressive cross-lingual representation in the encoder output space may deviate from the objective of ASR, although it is proven to benefit the ST and speech retrieval task.  

Lastly, we compare ZT-AED(asr) with ZT-AED and examine the effect of joint ASR-ST training. 
It is important to note that this comparison is conducted under a controlled setting where both ASR and ST data originate from the same source, which guarantees the joint ASR-ST training does not introduce new training data. 
In this setting, we surprisingly found that there is no ASR performance gain from additional ST training. 
We speculate that the empirical advantages of joint ASR-ST training may come from the new data sources introduced by ST.

\begin{figure}[h]
 \centering
 % Subplot 1 (left)
 \begin{subfigure}[t]{0.495\linewidth}
 \centering
 \includegraphics[width=\linewidth, height=4.7cm]{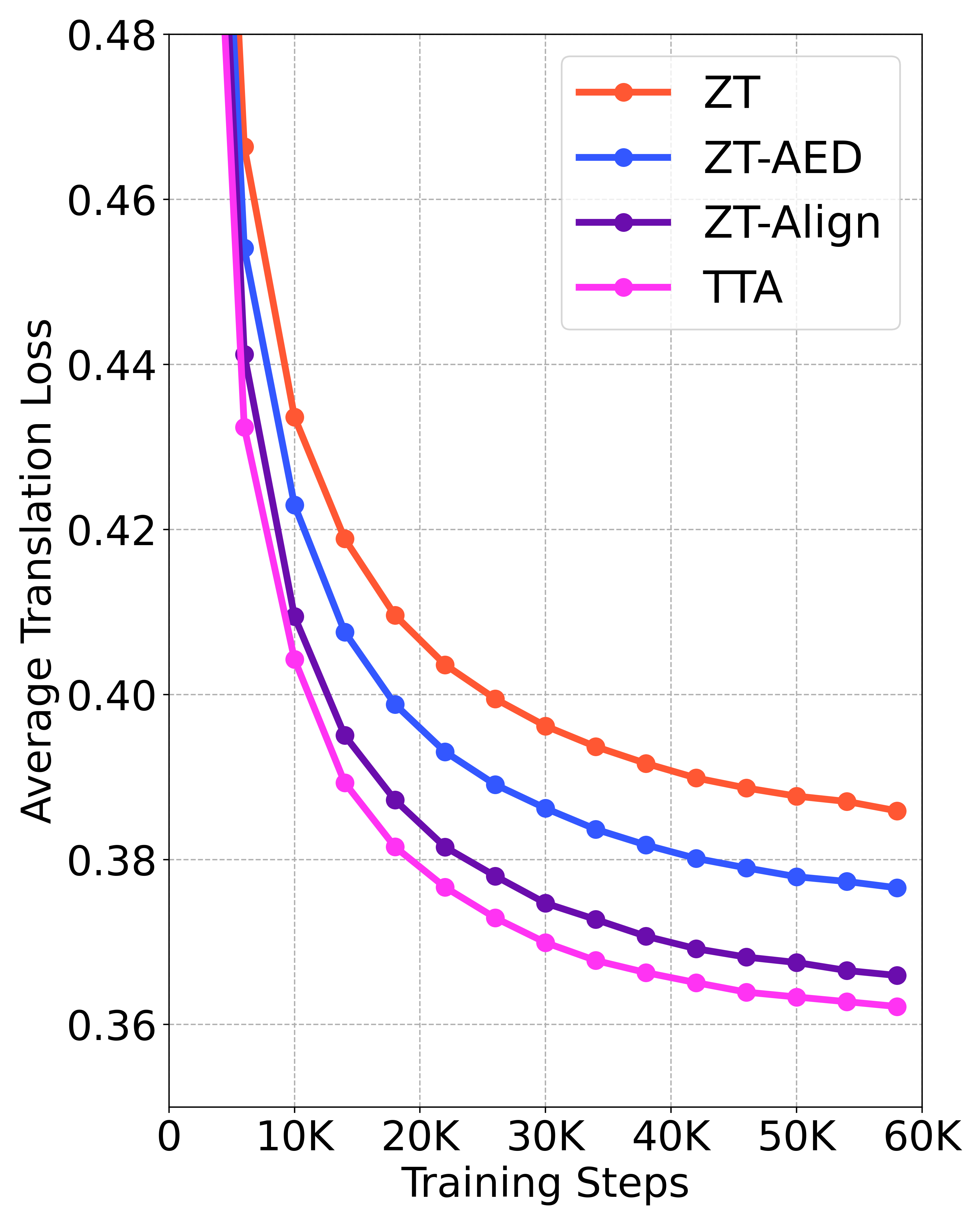}
 \caption{Speech translation probing.}
 \end{subfigure}
 % Subplot 2 (right)
 \begin{subfigure}[t]{0.495\linewidth}
 \centering
  \includegraphics[width=\linewidth, height=4.7cm]{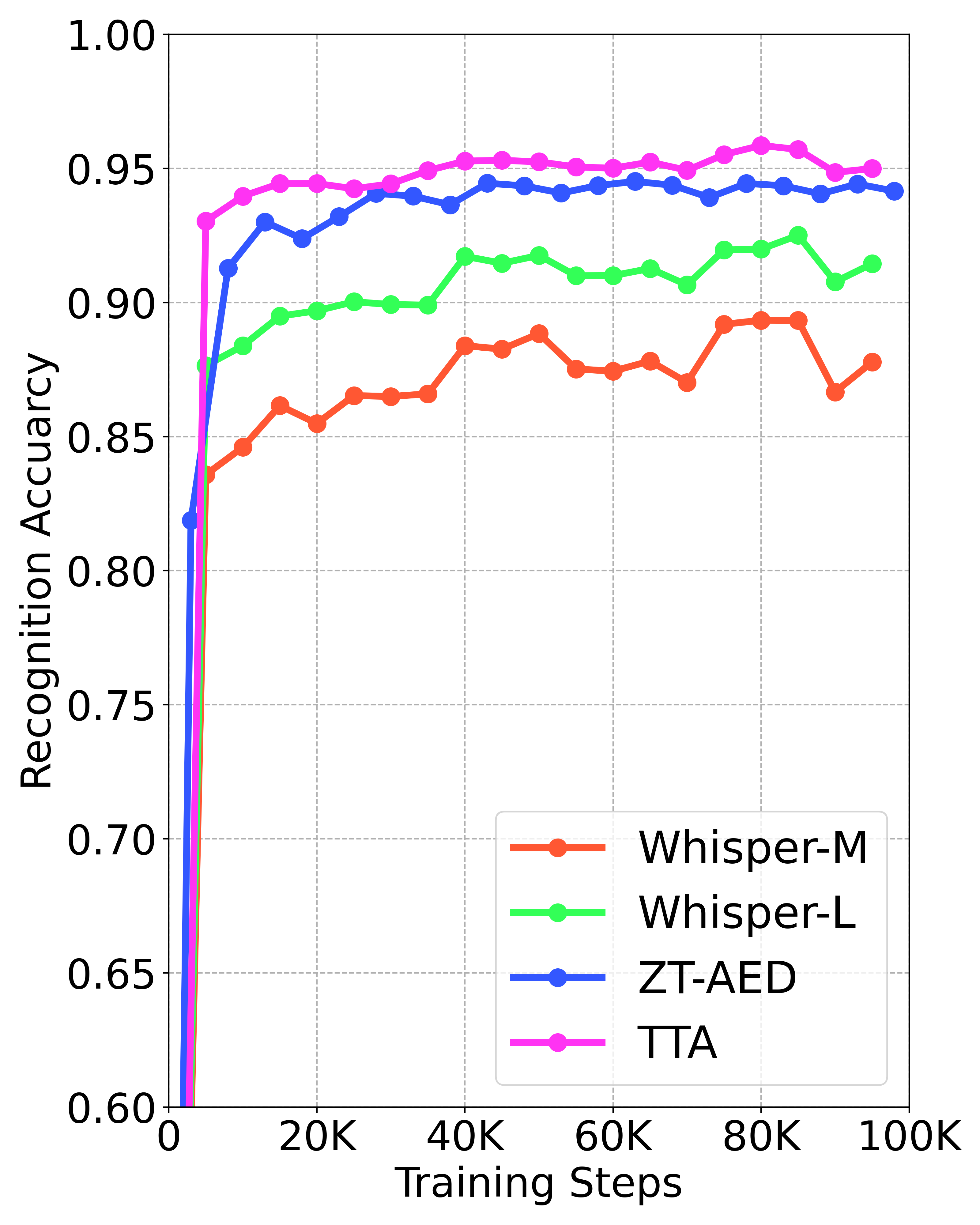}
    \caption{Training accuracy for ASR-LLM.}
  \end{subfigure}
  
  \caption{Two linear probing experiments. (a) compares averaged validation losses of the speech translation probing task with different encoders on CoVoSTv2. (b) shows the recognition accuracy curves, comparing various encoders in the ASR-LLM training using Aishell2 and Librispeech.} 
  \label{fig:training_dynamic}
  \vspace{-2mm}
\end{figure}

\begin{table}[h]
    \centering
    \caption{Recognition performance comparison of different encoders in ASR-LLM. CER (\%) for Aishell and WER (\%) for Librispeech.}
    \label{tab:asr-llm-wer}
    \resizebox{0.9\linewidth}{!}{
        \setlength{\tabcolsep}{4pt}
        \renewcommand{\arraystretch}{0.9}
        \begin{tabular}{l|cccc}
            \toprule
            ~ & Whisper-M & Whisper-L & ZT-AED & TTA \\
            \midrule
            Aishell & 5.47 & 4.87 & 2.92 & \textbf{1.92} \\
            Librispeech & 4.66 & 3.64 & 2.30 & \textbf{1.95} \\
            \bottomrule
        \end{tabular}
    }
    \vspace{-2mm}
\end{table}

\subsection{Semantic Encoder Evaluation in ASR-LLM}
To further evaluate the effectiveness of semantic representations learned by the TTA encoder, we employ ASR-LLM with \textit{Qwen2.5-7B-Instruct} as a linear probing task for comparative analysis. A single MLP layer is trained on top of each encoder to align its output representations with the token embedding space of LLM. 
The whole system is then trained on ASR using speech inputs and text-based prompts like \texttt{\small `Please repeat the following content:'}.
All models are trained under consistent settings using the Aishell and LibriSpeech datasets with a batch size of 4 for 100,000 steps.
The recognition performance is evaluated on Aishell-1 test-set and LibriSpeech test-clean.

As illustrated in Figure \ref{fig:training_dynamic}.(b), the training dynamics of ASR-LLM models with different encoders reveal that Whisper-based models demonstrate markedly lower optimization efficiency compared to those with Zipformer encoders.
% \textcolor{blue}{The WER results in Table \ref{tab:asr-llm-wer} indicate that, recognition performance on Chinese data generally improves under the ASR-LLM framework, whereas a slight degradation is observed for English data.} This discrepancy may be attributed to the inherent preference over Chinese of the Qwen LLM. 
% Notably, the ASR-LLM model with TTA encoder exhibits superior recognition performance comparable to the Transducer decoding from TTA model. 
The evaluation results are shown in Table \ref{tab:asr-llm-wer}. TTA series models clearly outperform Whisper series models in the ASR-LLM setting.
In contrast to the ZT-AED model, the TTA approach underscores the importance of explicit semantic alignment in effectively integrating speech representations with LLMs.
Additionally, we observe that the TTA model demonstrates significantly improved computational efficiency, operating at approximately twice the speed of Whisper Large-v2 and 1.5 times of Whisper Medium during training. 
% This efficiency gain is achieved despite the fact that the TTA model can indeed processes more samples within each batch.

\section{Conclusions}
In this work, we present TTA, an efficient Zipformer-based model capable of performing MASR and ST with high-quality semantic representations across ten languages.
With less than 300M parameters, TTA achieves superior performance in both speech recognition and translation tasks, while demonstrating higher efficiency and effectiveness within the downstream LLM integration task.
Through extensive experimental analysis, we systematically examine the mutual influence between MASR, ST, and cross-lingual representation learning, highlighting the importance of contrastive semantic alignment.
To promote reproducibility and further research, the model weights and codes will be released as part of the \textit{Auden} project.

{\footnotesize
\bibliographystyle{IEEEbib}
\bibliography{strings,refs}
}
\end{document}